# POWERING AN ECOSYSTEM OF PEDAGOGICAL AI AGENTS: A VALIDATION STRATEGY FOR A UNIFIED DATA ARCHITECTURE

**N. Theodora, P. Thajchayapong, A.K. Goel**

*Georgia Institute of Technology (UNITED STATES)*

## Abstract

The application of AI in education has evolved from monolithic intelligent tutoring systems to a diverse ecosystem of pedagogical agents, including conversational assistants, virtual coaches, and adaptive tutors. This shift requires a unified and scalable data architecture to manage the complex information feedback loops between human instructors, learners, and the varied AI agents. The design, development, and deployment of the data architecture in turn raises a critical issue of validation.

This paper addresses this critical need by describing a practical validation strategy for a high-volume data pipeline developed as part of a data architecture for AI-augmented adult learning at the National AI Institute for Adult Learning and Online Education. Our approach involves a two-stage testing methodology to ensure both functional diversity and real-world scalability. First, the QA environment uses a blend of synthetic and real-world data to validate functional correctness across various event types produced from learner and agent interactions. Following this, the production environment successfully processed a total of over 2.7 million production requests across 21 successful runs carrying authentic event data from a large-scale online program.

This validation process surfaced crucial insights into data privacy, a key challenge when handling varied data from multiple AI agent data sources. By outlining a replicable testing strategy for a unified data backbone, this research offers a clear framework for institutions and developers aiming to build and support their own heterogeneous suites of AI-powered learning tools.

## 1 INTRODUCTION

Artificial Intelligence (AI) in education is moving from single-purpose intelligent tutoring systems toward ecosystems of pedagogical agents, including conversational assistants, virtual coaches, social agents, and adaptive tutors. As these tools multiply, personalized and scalable learning depends on data architectures that can coordinate information flows among learners, instructors, platforms, and AI agents. This need is especially important in adult learning, where reskilling and upskilling require flexible support across heterogeneous environments. The National AI Institute for Adult Learning and Online Education (AI-ALOE; aialoe.org) addresses this challenge by developing AI techniques and tools for supporting adult learning at scale [1].

The Architecture for AI-Augmented Learning (A4L) provides the architectural foundation for this work [2]. Earlier implementations demonstrated the feasibility of connecting learning-management-system and AI-tool data, but they were constrained by fragmented sources, manual handling, and limited institutional integration. A4L 2.0 responds through a modular, cloud-native pipeline built around interoperability, software-defined infrastructure, and privacy-by-design [3]. It supports continuous data flows across learning management systems, student information systems, and AI-powered tools, thereby enabling analytics, dashboards, and feedback loops among learners, instructors, and agents.

This architectural direction is consistent with broader developments in modern data engineering, where disciplined data pipelines support scalable analytics and machine learning services [4]-[6]. In parallel, data observability extends monitoring from system uptime to data health across freshness, quality, volume, schema, and lineage [7]. Data contracts further emphasize explicit agreements about schema, reliability, and freshness between producers and consumers [4]. For AI-rich learning environments, these ideas are prerequisites for trustworthy personalization. The educational setting adds a distinctive validation problem. AI social agents such as Social Agent Mediated Interactions (SAMI) have been studied for fostering social presence and belonging in online classrooms [8], [9]. Their interaction streams include forum posts, introductory messages, and AI-mediated responses that can support fine-grained learning analytics. Yet these same streams contain semi-structured and unstructured text where personally identifiable information (PII) may appear unpredictably. Traditional example-based tests can miss such edge cases, motivating property-based testing approaches that generate broader sets of input cases from data specifications [10].

This paper presents an empirical validation of A4L 2.0 as a unified data architecture for pedagogical AI ecosystems. Rather than introducing a new architecture, the contribution is a practical validation strategy that tests whether the architecture can ingest, transform, pseudonymize, store, and monitor diverse educational events at scale. The study uses synthetic and authentic classroom-generated data, including learner and SAMI interactions, and evaluates the pipeline across QA and production-like environments.

# 2 METHODOLOGY

## 2.1 Unified data architecture and validation context

A4L 2.0 is a modular, cloud-native architecture for secure, scalable, and reusable educational data processing. Its design emphasizes interoperability through 1EdTech standards, repeatability through infrastructure as code, and privacy-by-design through pseudonymization and auditable event handling. The architecture spans ingestion, processing, analytics, and visualization, while the present study focuses on the Data Engine 2.0 layer responsible for accepting, validating, transforming, pseudonymizing, and storing educational events (Fig. 1).

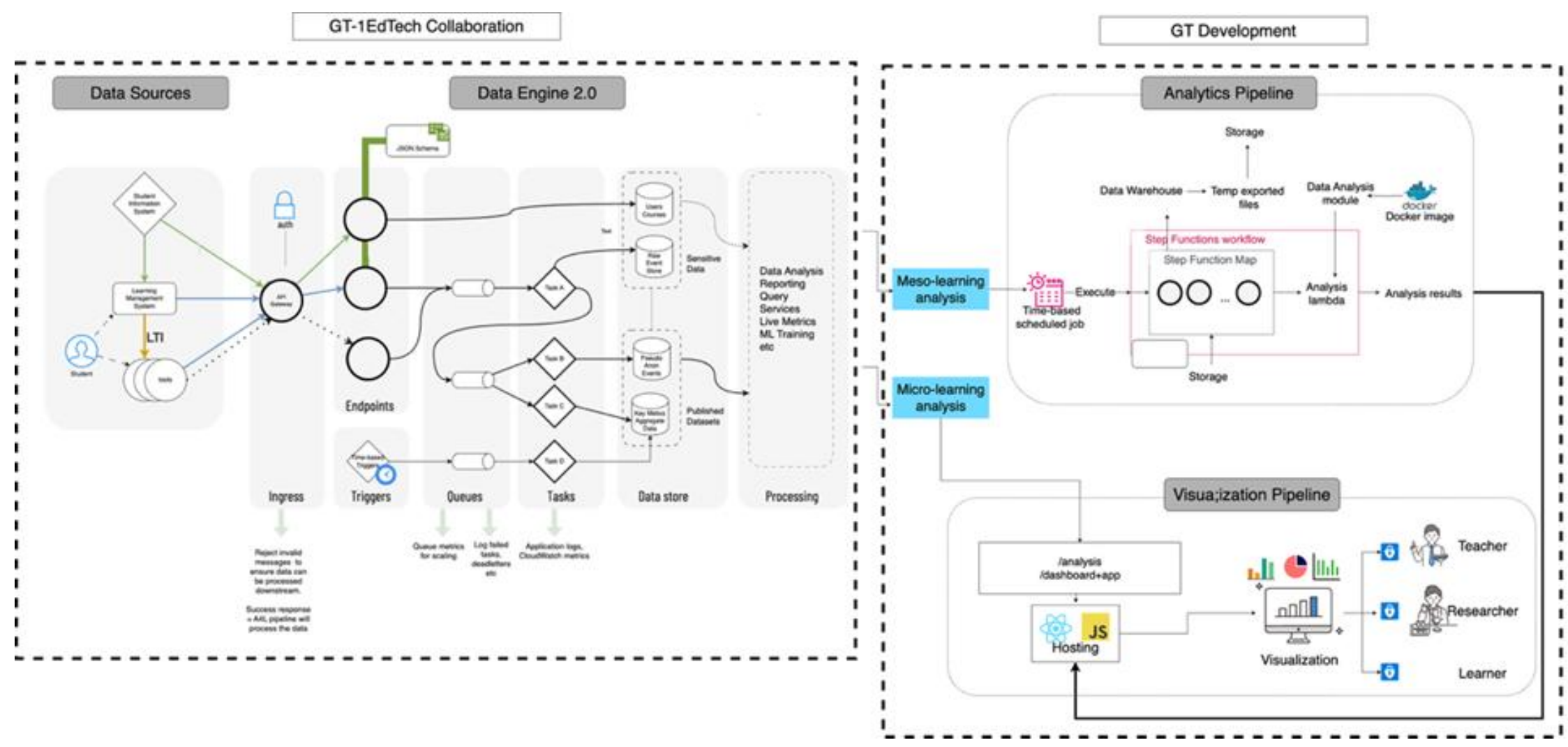

*Figure 1. Conceptual architecture of A4L showing the end-to-end pipeline: Data Engine 2.0 for ingestion and processing, Analytics Pipeline for scheduled learning analyses, and Visualization Pipeline for dashboards and AI-driven insights [2].*

From Fig. 1, the data engine follows an asynchronous event-driven workflow. Events generated by educational platforms and AI tools are sent through API endpoints using standards such as Caliper and Edu-API. Caliper Analytics is a 1EdTech standard for collecting and exchanging fine-grained learning activity and tool-use data from digital learning environments in a structured event format [11]. Edu-API is a complementary 1EdTech standard for exchanging core educational enterprise data, such as people, courses, offerings, organizations, and enrollments, so that activity events can be interpreted within their institutional and course context [12]. API Gateway triggers ingest Lambda functions that authenticate and validate incoming payloads before accepted events are placed into dedicated Amazon SQS queues, with dead-letter queues for failure isolation and replay. Downstream containerized AWS Lambda functions then perform transformation and pseudonymization before writing records into two Amazon RDS stores: a restricted sensitive datastore for auditability and a published datastore for pseudonymized analytics. AWS Cognito supports access control, and AWS CloudWatch provides component-level monitoring.

## 2.2 Validation design and data sources

The study was organized as a year-long validation effort spanning Spring, Summer, and Fall 2025 terms at Georgia Institute of Technology. The submission abstract frames the work as a two-stage method: QA validation followed by production validation. Operationally, the internal testing plan unfolded in three phases. The Spring term established baseline functionality with synthetic Caliper Analytics version 1.2 (v1p2) events, where v1p2 refers to the 1EdTech Caliper 1.2 specification used to represent learning activity data as standardized event records [11]. This version label fixes the expected Caliper event vocabulary, profiles, identifiers, and timestamp formatting, allowing the team to test schema compliance and end-to-end ingestion before introducing authentic KBAI classroom datasets.The Summer term introduced authentic forum activity data from the Summer 2024 Georgia Tech class on Knowledge-Based AI [13] and added performance testing. The Fall term expanded testing to additional authentic datasets, including introductory posts and SAMI post-response interactions, and moved the pipeline into production-like deployment.

The validation posture was black-box and architecture-based. Following Myers et al. [14], the tests evaluated behavior against external expectations rather than internal code paths. Following Architecture-Based Testing, the focus was conformance of the implementation to its intended architectural design [15]. This posture mirrors practical use: events enter through API Gateway and are validated through their downstream appearance in sensitive and published RDS stores.

The data included both synthetic and authentic event streams. Synthetic data covered more than forty event types and was used for early functional and integration testing. Authentic data came from the Summer 2024 KBAI class and included forum activity, introductory posts, and SAMI interaction records. Forum activity and introductory posts were modeled as Caliper MessageEvent records. In Caliper, a MessageEvent represents a person posting a message or marking a post as read or unread, making it appropriate for message-based learner activity [11]. SAMI post-response data were modeled as Caliper QuestionnaireItemEvent records; this event type captures a respondent starting, completing, or skipping an individual questionnaire item, and a completed item may include a generated response, allowing the learner post to be represented as the questionnaire item and the SAMI output as the response [11].

The real KBAI datasets were converted into Caliper-compliant JSON payloads using a configuration-driven generator with mapping CSV files and JSON templates. Before payloads were sent, the generator performed pre-send conformance checks for IRI prefixes, URN-UUIDv4 identifiers, and ISO-8601 UTC timestamps. IRI prefixes identify Caliper entities and resources using valid web-style identifiers, URN-UUIDv4 identifiers provide globally unique event IDs in the form urn:uuid:<UUID>, and ISO-8601 UTC timestamps standardize event times in a consistent machine-readable format such as YYYY-MM-DDTHH:mm:ss.SSSZ; these checks helped detect malformed payloads before API ingestion [11].

## 2.3 Validation procedures and observability

Functional validation examined three paths. First, ingress checks expected valid Caliper events to return HTTP 200 responses and invalid events to return HTTP 400 responses. Second, RDS checks tested event ID uniqueness, row-count parity, primary-key alignment, type/version consistency, and appropriate obfuscation across sensitive and published databases. Third, ingress-to-RDS checks compared transmitted JSON payloads against stored records in the sensitive database, supporting end-to-end inference from API ingestion through downstream storage. Postman was used as the canonical sender, with Python requests as a secondary cross-check.

Performance validation used an automated API testing tool (Postman Runner) to simulate ramp-up and peak workloads with varying virtual-user levels. In Fall 2025, the target average response-time threshold was tightened to 300 ms. Envelope-size experiments under 20 virtual users were used to select 10 events per envelope for production-like testing, because this was the largest tested envelope size that kept average response time below the threshold. Production drills included write-only, read-only, and mixed read/write workloads.

Observability was treated as part of validation. CloudWatch dashboards monitored API Gateway, Lambda, SQS, dead-letter queues, and RDS, with dashboard windows aligned to Postman run timestamps. The study also maintained a master test tracker, GitHub issues, Postman reports, CloudWatch dashboard snapshots, JUnit XML outputs, and CSV exports. This documentation supported traceability and made the validation strategy reproducible.

# 3 RESULTS

## 3.1 Functional validation

Functional results showed strong core performance across both QA and production. At the application-test level, QA produced 37 passes, 20 partial passes, and 12 failures; production produced 39 passes, 20 partial passes, and 10 failures. Integration tests showed no outright failures in either environment, while backward-compatibility checks passed in both environments. These results indicate that the pipeline consistently handled core ingestion and storage workflows, while a smaller set of tests exposed specific data-consistency problems.

Sample-fixture testing in QA confirmed that valid schemas were accepted through both Postman and Python. Event ID uniqueness, row-count parity, and primary-key checks also passed. The main failures appeared when comparing sensitive and published records on primary key and event ID, in pseudonymization checks, and in sensitive-to-published relation checks. The authentic KBAI datasets produced the same pattern in both QA and production: API submission, event uniqueness, row-count parity, pseudonymization, row-type consistency, and SAMI transmission checks passed, while sensitive-to-published event parity and relation checks repeatedly failed. The stability of these failures suggests persistent transformation or pseudonymization issues rather than environment-specific instability.

## 3.2 Performance and scalability

Performance testing demonstrated that the architecture could sustain realistic high-volume educational traffic. Successful QA KBAI runs with 20 virtual users generally achieved average response times between approximately 216 and 277 ms, with throughput ranging from about 20 to 47 requests per second across the main successful KBAI runs. Failed QA runs were concentrated in deterministic or larger-batch scenarios where response times exceeded the 300 ms threshold.

Production testing showed similar or stronger behavior. Nearly all 60-minute production runs with 20 virtual users passed. Successful runs typically processed approximately 113,000 to 154,000 total requests per run, sustained 31.24 to 42.67 requests per second, and maintained average response times between 221 and 291 ms. Aggregating the successful production runs (21 runs) yielded 2,738,521 total requests carrying authentic KBAI data. One clearly failed production run occurred under a slow client-network condition, with throughput dropping to 12.7 requests per second and average response time rising to 1,013 ms.

The representative production run shown in Fig. 2 illustrates this production pattern in a single ramp-up cycle. Run #77 processed 136,873 requests, sustained 37.85 requests per second, maintained an average response time of 253 ms. This run is useful because it shows the load profile directly: virtual users increased during the window while response times remained within the Fall 2025 target range.

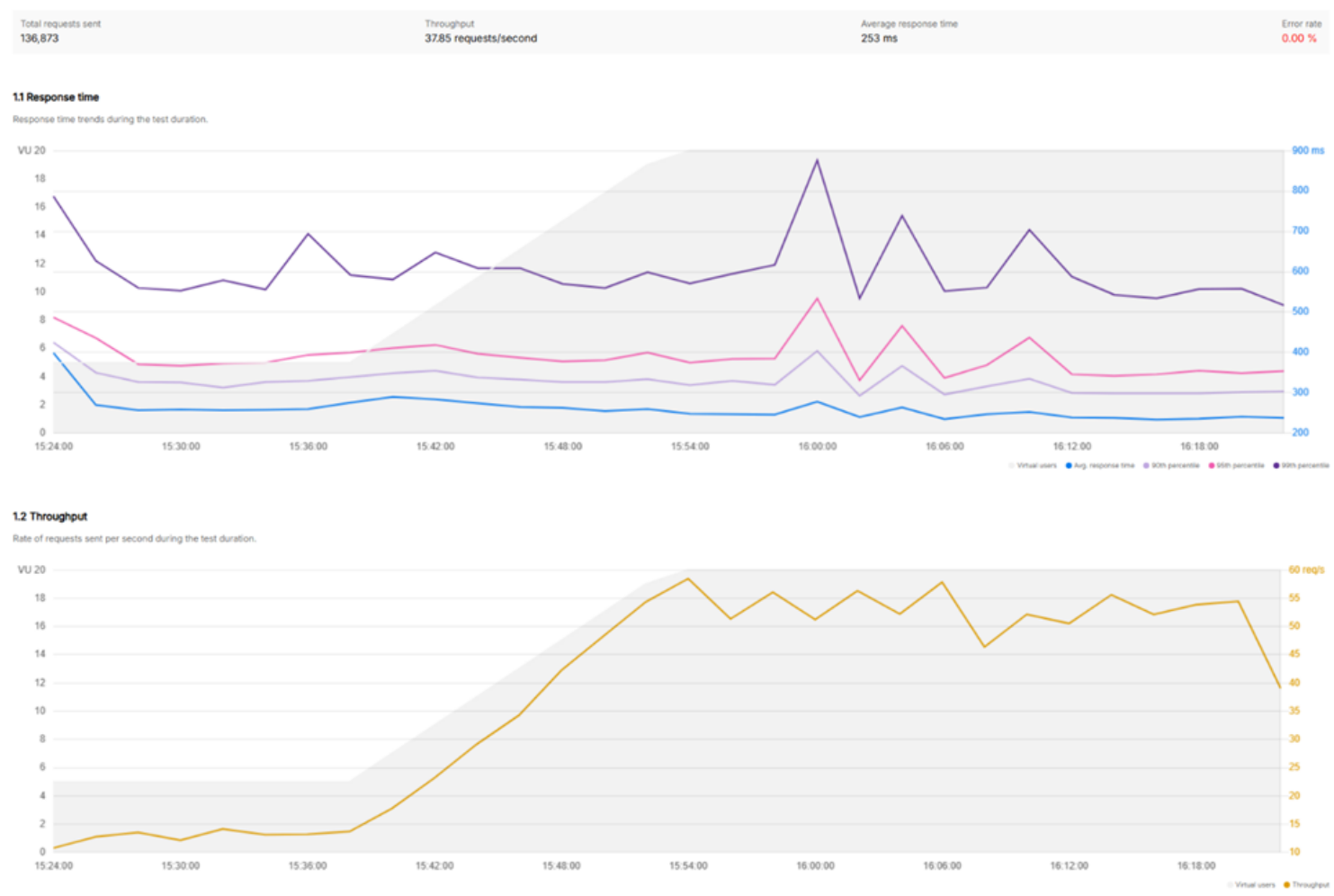

*Figure 2. Postman production load test run #77 for the AI-ALOE Reference Pipeline.*

The production total reported above therefore reflects repeated successful high-volume delivery cycles rather than a single isolated stress test. Because the Fall production tests used batched Caliper envelopes, the analysis emphasizes accepted delivery submissions, sustained ingress behavior, and stable response times across repeated runs using authentic KBAI data.

Component-level monitoring helped interpret these results. API Gateway showed burst-aligned request spikes with rare 4XX or 5XX errors. Lambda invocations and concurrency scaled with load, with minimal errors, rare throttling, and only brief duration spikes. SQS sent, received, and deleted counts moved largely in lockstep, indicating healthy queue flow. Dead-letter queue activity remained minimal. RDS showed the clearest pressure under sustained load: CPU utilization, connection counts, write throughput, IOPS, and disk queue depth rose during ingestion peaks and then recovered. In one-day production analysis, database CPU repeatedly reached 95-97%, but the system remained functionally sustainable. Fig. 3 provides a representative RDS dashboard view of this pattern, showing intermittent CPU and I/O spikes, fluctuating database connections, and sustained low freeable memory during production load testing.

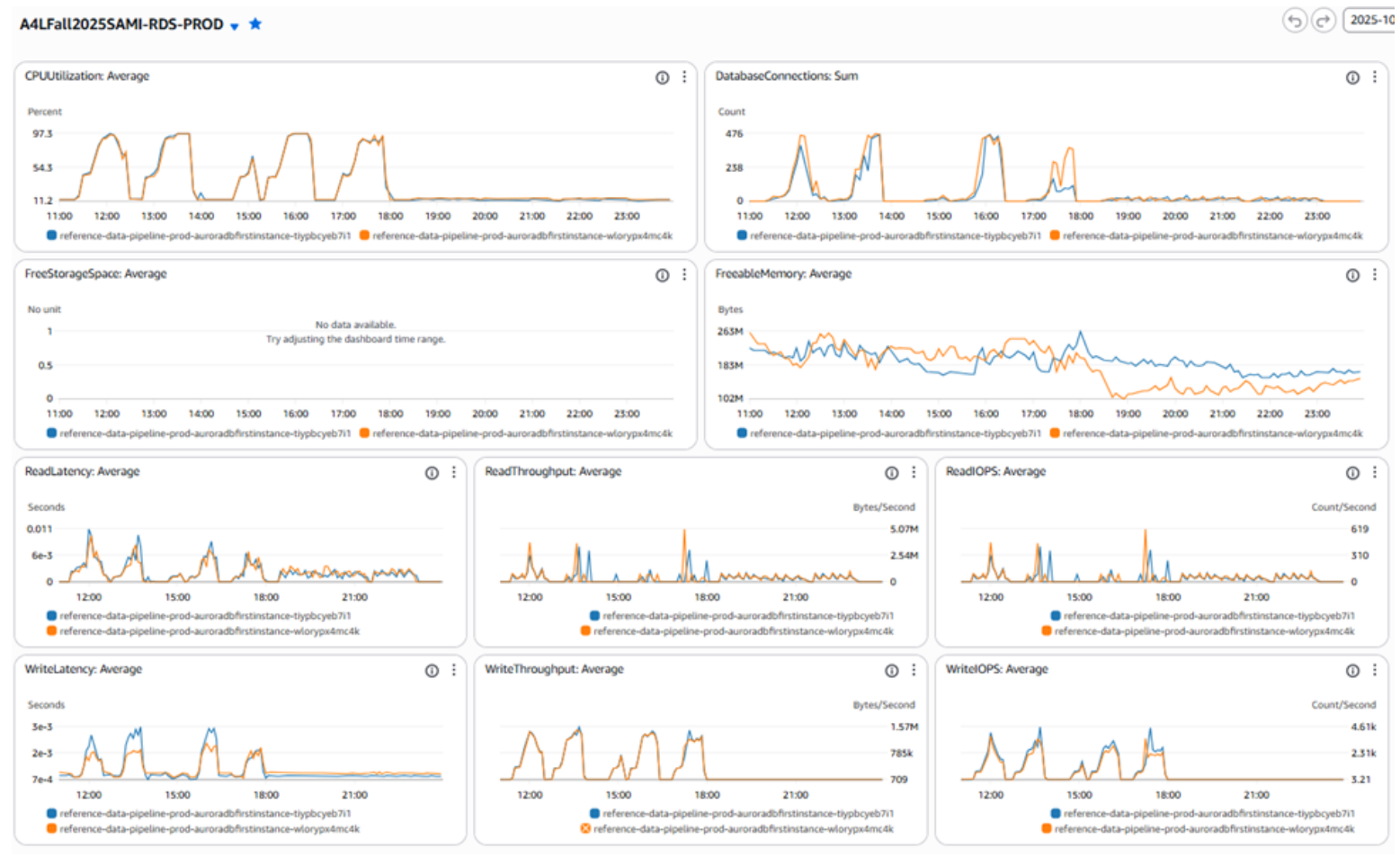

*Figure 3. Composite RDS metrics for the sensitive and published read-write datastores during production load testing, showing intermittent CPU and I/O spikes, fluctuating database connections, and sustained low freeable memory.*

The degraded-network production run helped separate client-side effects from pipeline-side behavior. When local connectivity was slow, Postman recorded lower throughput and higher response time before requests reached API Gateway. After successful ingress, AWS-side dashboards did not show comparable increases in persistent errors, queue aging, or dead-letter traffic, suggesting that the asynchronous AWS path continued to process events normally once messages entered the pipeline.

*Table 1. Summary of validation outcomes across QA and production environments.*

| Validation area | QA outcome | Production outcome | Main issue surfaced |
|---|---|---|---|
| Functional tests | 37 pass, 20 partial, 12 fail in application tests; valid-schema fixtures passed | 39 pass, 20 partial, 10 fail in application tests | Sensitive-to-published event parity and relation checks failed repeatedly |
| Authentic KBAI datasets | API submission, uniqueness, row counts, pseudonymization, and SAMI transmission passed | Same pattern as QA | Transformation and pseudonymization inconsistencies persisted |
| Performance/load tests | Main successful KBAI runs: approx. 216-277 ms average response | Main successful runs: 221-291 ms; 31-42 requests/s; 2,738,521 aggregate requests | RDS showed peak CPU, connection, and write-latency pressure |

| Validation area | QA outcome | Production outcome | Main issue surfaced |
|---|---|---|---|
| | time; approx. 20-47 requests/s | | |
| Infrastructure checks | SQS redrive and RDS schema checks passed; QA retention and multi-AZ failed by design for cost | SQS redrive, RDS retention, multi-AZ, and schema checks passed | Alarms and Lambda reserved concurrency were incomplete in both environments |

## 3.3 Privacy, schema, and infrastructure issues

The validation surfaced four recurring issue classes. First, data consistency between sensitive and published databases was the most frequent problem. Event IDs in published records were sometimes pseudonymized even when they were not PII, breaking referential integrity for rows sharing the same primary key. Additional divergence appeared in actor.id, @context, and extensions fields.

Second, schema validation was not always enforced strictly at the edge. Some malformed events with incorrect IRI or URN formats, misnamed properties, missing fields, or invalid data types were accepted instead of being rejected with the expected HTTP 400 response. Third, privacy validation showed both over-pseudonymization and under-pseudonymization. Regex-based pseudonymization can protect structured fields but cannot reliably detect names or other sensitive information embedded unpredictably in free text. Fourth, infrastructure validation showed that core data-path safeguards were present, but operational safeguards were incomplete: CloudWatch alarms were insufficiently configured, reserved concurrency was missing for Lambda functions, and CloudWatch logs were difficult to correlate manually at high event volumes.

Formal infrastructure checks reinforced this distinction between core reliability and operational readiness. SQS redrive policies and RDS table/schema checks passed in both QA and production, and the production clusters met backup-retention and multi-AZ expectations. QA intentionally lacked some redundancy controls for cost reasons. However, alarm configuration and Lambda reserved concurrency failed in both environments, indicating that the main data path was stronger than the surrounding automated operational safeguards.

# 4 CONCLUSIONS

## 4.1 Discussion

The results provide preliminary support for the central claim that A4L 2.0 can operate as a unified data backbone for heterogeneous learner- and agent-generated event streams. Across QA and production-like environments, the pipeline demonstrated reliable ingestion, storage, and monitoring behavior under realistic classroom and load conditions. This is important because much prior work on educational data architecture emphasizes design principles, while fewer studies validate real systems under production-like classroom conditions.

At the same time, the repeated sensitive-to-published failures show that functional ingestion is not enough. A privacy-preserving architecture must maintain semantic consistency across transformation stages. The observed event-ID mismatches and content divergence point toward the need for more formal governance, including clearer pseudonymization policies and data contracts that define which fields may change and which must remain stable for auditability [4].

The performance results also clarify the difference between scalability and production readiness. API Gateway, Lambda, and SQS generally scaled smoothly, and the pipeline maintained target latency and throughput in successful production runs. However, RDS pressure, missing alarm coverage, and absent Lambda reserved concurrency show that operational safeguards still require strengthening. In data-observability terms, the system has useful signals for latency, volume, backlog, and database load, but those signals need to be coupled with automated alarms and service-level objectives to reduce detection and resolution time [7].

The most important unresolved challenge is privacy validation for unstructured educational text. Current automated checks can verify known structured fields, but forum posts, introductory messages, and SAMI-related responses may contain PII in unexpected forms. This limitation is consistent with broader work showing that PII labeling and obfuscation in educational text often requires more advanced tools and human oversight [16]. Similar challenges have appeared in other large-scale learning infrastructures, including Total Learning Architecture experiments [18].

The study therefore argues for validation as an integral part of the data lifecycle. A shift-left approach should embed functional tests, privacy checks, schema enforcement, infrastructure validation, and observability earlier in development and deployment [17]. Automated release pipelines can also reduce configuration drift across environments [17]. The methodological value of this paper lies in demonstrating a replicable validation pattern: phased black-box testing, authentic classroom data, batched load testing, time-aligned CloudWatch monitoring, and explicit issue tracking.

## 4.2 Limitations and Future Directions

The study has several limitations. First, the black-box posture provided strong evidence about end-to-end behavior, but it limited fault isolation when failures appeared between ingestion, transformation, pseudonymization, and storage. Second, the authentic datasets came mainly from a single course context and emphasized MessageEvent and QuestionnaireItemEvent records, so future deployments should validate a wider range of event types, tools, courses, and institutional delivery patterns.

Performance coverage also remains bounded by the load-generation setup. Postman-based runs with local desktop resources were valuable for repeatable ramp-up and peak scenarios, but they cannot fully reproduce the diversity, burstiness, or network paths of a large institutional deployment. Several infrastructure checks still required manual inspection, and high-volume audit tracing through CloudWatch remained difficult when many events were submitted simultaneously.

Future validation should therefore move toward CI/CD-driven testing in which schema checks, privacy validation, infrastructure configuration checks, and operational alarms run automatically with each deployment [17]. Privacy validation should also extend beyond regular expressions by incorporating configurable pseudonymization policies and tools for detecting personally identifiable information in unstructured educational text [16]. For schema quality, source-system data contracts and property-based testing can help generate a broader set of edge cases before malformed events reach the production pipeline [4], [10].

## 4.3 Summary

This study presented an empirical validation of A4L 2.0 as a unified, cloud-native data architecture for AI-augmented learning. Using synthetic and authentic classroom-generated data, the study evaluated the pipeline across QA and production-like environments, focusing on functional correctness, scalability, observability, and privacy-related limitations. The findings show strong overall functional and scalability performance. Most core validation checks passed across QA and production, and production tests sustained average response times of roughly 221–291 ms and throughput of 31.24–42.67 requests per second across successful runs. The validation also demonstrated high-volume event delivery, with successful production runs aggregating more than 2.7 million total requests carrying authentic KBAI data.

The same process surfaced important unresolved challenges. Pseudonymization rules sometimes altered non-sensitive identifiers, schema validation did not reject all malformed events, observability safeguards were incomplete, and free-text learner-agent interactions remained difficult to validate automatically for PII. These findings show that trustworthy educational AI infrastructure depends not only on throughput, but also on data governance, traceability, and privacy assurance. Along with the Total Learning Architecture [18], A4L2.0 points to a new direction in data architectures for supporting AI-augmented learning at scale.